\documentclass[aps,showpacs,superscriptaddress,amsmath,amssymb,preprint]{revtex4}
\usepackage{graphicx} 
\usepackage{amsmath}
\usepackage{xcolor}
\usepackage{xcolor}
\usepackage{array}
\usepackage{multirow}  
\usepackage{caption}   
\begin{document}
\title{Dependence of angular momentum of fission fragments on total kinetic energy in spontaneous fission of $^{252}$Cf}

\author{A. Rahmatinejad}
\affiliation{Joint Institute for Nuclear Research, Dubna, 141980, Russia}
\author{T. M. Shneidman}\email{shneyd@theor.jinr.ru}
\affiliation{Joint Institute for Nuclear Research, Dubna, 141980, Russia}
\author{G.G. Adamian}
\affiliation{Joint Institute for Nuclear Research, Dubna, 141980, Russia}
\author{N. V. Antonenko}
\affiliation{Joint Institute for Nuclear Research, Dubna, 141980, Russia}
\affiliation{Tomsk Polytechnic University, Tomsk, 634050, Russia}

\begin{abstract}
A weak dependence of the angular momentum of $^{144}$Ba fragments, produced in the spontaneous fission of $^{252}$Cf, on the total kinetic energy (TKE) was recently observed \cite{Giha2025}. To investigate this phenomenon, we propose a model describing the evolution of the fissioning nucleus toward scission. The model assumes that after tunneling through the fission barrier, the nucleus can be represented by a superposition of dinuclear systems (DNS). We calculate main fission observables, including mass, TKE, and neutron multiplicity distributions, and compare them with experimental data. The angular motion in the DNS is treated quantum-mechanically, yielding the angular momentum distribution of DNS nuclei at scission configurations leading to $^{144}$Ba fragments. Within this framework, we successfully reproduce and explain the experimentally observed dependence of the average angular momentum of $^{144}$Ba on TKE.
\end{abstract}


\maketitle

\section{Introduction}

 Despite the long history of studying spontaneous fission (SF), new experimental data are still emerging, reviving interest in this subject.  Although there is a general understanding of key properties of SF process, various theoretical models are based on different assumptions to describe fission observables. New experimental findings not only validate and discriminate these models, but also offer fresh insights helping to refine our understanding of the fission mechanism.

Theoretical models of SF diverge particularly in two aspects. First, they differ in understanding at what elongation of fissioning nucleus the distribution of primary fission fragments is formed.
Second, an important question is how the available energy at this elongation is distributed between the excitation and  deformation energy of the fragments. Both aspects are crucial for describing the gross properties of SF, such as mass/charge distributions, total kinetic energy (TKE), and neutron multiplicity. However, the splitting between excitation and deformation energies appears particularly decisive in describing correlations between observables, such as the relationship between the average number of neutrons emitted from fission fragments and TKE, or between TKE and fission fragment mass \cite{Gook2014}.

The experimental data related to the measurement of the angular momenta of fission fragments,  have shown that fission fragments possess significant angular momentum, even in the SF of even-even nuclei that occurs from the $0^+$ state \cite{Wilhelmy1972,Pleasonton1972,Wolf1976,Akopian1994}. Mechanisms based on angular motion in scission configurations have been proposed in Refs.~\cite{Nix1965,Rasmussen1969,Zielinska1974,
Kadmensky2015,Randrup2021,Randrup2022,Bertsch2019,Marevic2021,Bulgac2022,Shneidman2003,Dossing2024,Scamps2023} and successfully applied to describe the average angular momenta of fission fragments.
The measurements revealed an absence of correlation between the spins of conjugate fragments \cite{Wilson2021}. Another interesting observation is related to the strong correlation (saw-tooth behavior) between the average angular momentum of fragments and their mass. These observations provide a good ground for testing existing SF models.

In different models, the absence of correlation between angular momenta of the fission fragments is attributed to the large moment of inertia associated with their relative rotation \cite{Scamps2023,Dossing2024,Shneidman2024}. However, there are two different approaches to explain the sawtooth behavior of average angular momentum as a function of fragment mass.
In Ref.~\cite{Randrup2022}, it is assumed that most of the energy available at scission contributes to excitation energy of the fragments while fragments deformations remain close to those of their ground states. In this case, the sawtooth behavior arises from different structure of excitation spectrum of collective states for conjugate fragments.   Conversely, in Refs.~\cite{Shneidman2024, Marevic2021}  most of the energy is stored as the deformation energy of the fragments, with relatively low excitation energy. Here, the saw-tooth pattern originates from the fact that the softer fragment possess larger deformation resulting in larger average angular momentum.

Since both approaches describe sawtooth behavior, the question of the balance between excitation and deformation energy at scission might require examining correlations among other observables. Recently, such a correlation was reported in Ref.~\cite{Giha2025} for the fission fragment $^{144}$Ba, where almost an absence of correlation between its angular momentum and TKE was observed. This new data supplemented by the correlation between average number of the emitted neutrons and average TKE \cite{Gook2014} gives a possibility to answer this question.

 At first glance, in contrast to the experiment, both types of model should lead to an increase of the average angular momenta of the fragments with decreasing TKE.
 If the fissiong nucleus at scission is hot then angular momentum of the fission fragments is roughly described by the Boltzmann distribution \cite{Randrup2022}.  Therefore, the decrease of TKE and subsequent increase of excitation energy would lead to an increase of angular momentum of fission fragments.
 If available energy is mainly stored as deformation energy, the decrease of TKE would lead to increase of deformation as shown in Ref.~\cite{Shneidman2024}, that  can also lead to the increase of angular momentum. Therefore, this puzzling behavior of angular momentum vs TKE is a good test for the models. It is important, in addition to the correlations, to describe simultaneously other fission observables within unique approach.

 For this purpose we employ the fission model introduced in Ref.~\cite{Rahmatinejad2024_2} supplemented by the description of collective states related to the angular motion of fission fragments at scission \cite{Shneidman2024}. As assumed in this model, after tunneling through the fission barrier the nucleus can be represented by a superposition of two fragments at touching.
 Variations in deformation and mass asymmetry occur in these systems until dinuclear systems (DNS) decay.  The characteristics of DNS fragments, including their deformations, mass asymmetry, and energy, play a decisive role in shaping the observable distributions associated with fission.
 The proposed approach is a modification of the scission point model which takes into account that not all scission configurations are  dynamically equally  achievable. It can be thought of as a model of random walk among various scission configurations. Recent microscopic calculations in the framework of time-dependent density functional theory   support the scission point model \cite{Bulgac2019}.

 The paper is organized as follows. In Sec.~\ref{model}, we present the model used to describe the SF process as an evolution of initially formed DNS configurations toward scission. The basic assumptions of the model and the main characteristics defining different DNS configurations—including potential energies, excitation energy, and level densities—are detailed in Sec.~\ref{model1}. Section~\ref{model2} introduces the method for calculating the primary fission fragment distribution. Section.~\ref{Sec.I} provides a quantum-mechanical treatment of collective angular motion in the DNS. Finally, Sec.~\ref{model3} presents the expressions used to derive various primary fission fragment distributions, such as mass, TKE, neutron multiplicity, and average angular momentum.
The calculated fission observables for the SF of $^{252}$Cf are presented in Sec.~\ref{result1}. In Sec.~\ref{result2}, we identify the DNS configurations leading to the formation of the post-scission fragment $^{144}$Ba and in Sec.~\ref{result3} we analyze angular vibrations in the selected DNS. Section~\ref{result4} presents the calculated average angular momenta of primary fission fragments as a function of TKE. Finally, Sec.~\ref{result5} discusses the average angular momentum of post-scission $^{144}$Ba fragments as a function of TKE. The conclusions are summarized in Sec.~\ref{conclusion}.

\section{Model \label{model}}

\subsection{Fissioning nucleus at scission \label{model1}}

After crossing the fission barrier, the fissioning nucleus $^AZ$  is treated as a superposition of various DNS. As a DNS we understand a system of two fragments ($i=1,2$) in touching configuration characterized by their charges ($Z_{1,2}$) and masses ($A_{1,2}$), where $Z_{1}+Z_{2}=Z$ and $A_{1}+A_{2}=A$. The shape of each fragment is modeled as an axially-symmetric ellipsoid with major-to-minor axes ratio denoted as $b_{1,2}$. The quantities $b_{1,2}$ are related to the conventional quadrupole deformation parameters $\beta_{1,2}$ by equating the quadrupole moments calculated for ellipsoidal and  spheroidal shapes:
\begin{eqnarray} \label{eq0}
    Q(\beta_{1,2})=Q(b_{1,2}),\nonumber \\
b_{1,2} \approx\frac{1+\sqrt{5/4\pi}\beta_{1,2}}{1-\sqrt{5/16\pi}\beta_{1,2}}.
\end{eqnarray}
In the following, each DNS is characterized by a set ($Z_{i},A_{i},\beta_{i}, i=1,2)$.

Due to the competition between attractive nuclear and repulsive Coulomb parts of nucleus-nucleus interaction, the potential energy of DNS as a function of relative distance coordinate $R$ has a pocket or at least is flat in the vicinity of touching configuration at $R=R_m(\beta_1,\beta_2)$. Therefore, the DNS does not decay instantly. Instead, the decay of DNS competes with an evolution in mass/charge asymmetry coordinates and in deformation of the fragments.

The shapes of the DNS nuclei, their masses, charges, and the distribution of angular momenta, as well as the excitation energy of DNS at the moment of decay determine the primary fission fragment distributions.

For each DNS one can calculate the potential energy $U(Z_{1,2},A_{1,2},\beta_{1,2})$ and excitation energy $E^{*}(Z_{1,2},A_{1,2},\beta_{1,2})$. The potential energy of DNS consists of terms describing energies of the constituent fragments $(U_{i}^{LD}+\delta E_{i}^{sh})$ and the energy of their interaction $V_{int}$ \cite{Adamian1996}:
\begin{eqnarray}\label{ebin}
U(Z_{1,2},A_{1,2},\beta_{1,2})&=&
U_{1}^{LD}(Z_{1},A_{1},\beta_{1})+\delta E_{1}^{sh}(Z_{1},A_{1},\beta_{1})\nonumber\\
&+&U_{2}^{LD}(Z_{2},A_{2},\beta_{2})+\delta E_{2}^{sh}(Z_{2},A_{2},\beta_{2})\nonumber\\
&+&V_{int}(Z_{1,2},A_{1,2},\beta_{1,2},R_m(\beta_1,\beta_2)).
\end{eqnarray}
The energy of the $i$-th fragment consists of the liquid-drop part $U_{i}^{LD}$ and shell correction $\delta E_{i}^{sh}$.
The liquid-drop energies are calculated within the prescription given in Ref.~\cite{Myers1966}. Shell corrections are acquired using Strutinsky approach with the single-particle spectra of deformed Nilsson-like Hamiltonian \cite{Maruhn1972}. The parameters of spin-orbit interaction are taken as in Ref.~\cite{Ragnarsson1995,Kuzmina2012}.

The interaction potential $V_{int}$ in Eq.~\eqref{ebin} is calculated for the pole-to-pole touching configuration of the fragments. The details of calculations are presented in Refs.~\cite{Adamian1996,Shneidman2024}. It is assumed that energies of  zero-point vibrations in relative distance and in the orientation of fragments are taken into account by the choice of  effective interaction. The orientation dependence of DNS potential energy is excluded from $V_{int}$. The collective angular motion of the fragments is  treated as an intrinsic excitations of DNS, therefore this dependence contributes to the energies of collective angular motion in DNS  (see Sec.\ref{Sec.I}).

An important ingredient needed to characterize a DNS is the excitation energy. Denoting the energy of the fissioning nucleus as $U_{0}(Z,A)$, the excitation energy is given as
\begin{eqnarray}
E^{*}(Z_{1,2},A_{1,2},\beta_{1,2})&=&U_{0}(Z,A)-U(Z_{1,2},A_{1,2},\beta_{1,2}),\nonumber \\
U_{0}(Z,A)&=&U^{LD}(Z,A,\beta^{(gs)})+\delta E^{sh}(Z,A,\beta^{(gs)}),
\label{Eq13}
\end{eqnarray}
where the ground-state deformations $\beta^{(gs)}$ of the $^{252}$Cf are taken from \cite{Moller2016}.

The intrinsic state of the DNS is characterized by its level density $\rho(E^{*},DNS)$, which consists of intrinsic level density and the level density of collective states associated with the angular motions in DNS.
The intrinsic level density of DNS is calculated as a folding of the level densities $\rho_{1,2}$ of its constituent fragments \cite{Bezbakh2015}
\begin{eqnarray}\label{M6}
\rho_{int}(E^{*},DNS)=\int\rho_{1}(\varepsilon)\rho_{2}(E^{*}-\varepsilon){\text{d}}\varepsilon.
\end{eqnarray}
To obtain the level densities of DNS fragments, the superfluid formalism is used \cite{Dekowski1968,Adeev1975,Behkami1973}.
The required single-particle spectra are calculated by diagonalization of the Nilsson Hamiltonian for given mass, charge, and deformation of the fragment. This approach allows us to take into account not only the excitation energy but also the deformation dependence of the level densities \cite{Rahmatinejad2020,Rahmatinejad2021,Azam_ratios_2022, Rahmatinejad2024}.

The integrant in Eq.~\eqref{M6} is strongly peaked function of $\varepsilon$, so the excitation energies of DNS fragments $E^{*}_{1}$ and $E_{2}^{*}=E^{*}-E^{*}_{1}$ can be obtained from the condition
\begin{eqnarray}\label{M6b}
\left.\left(\frac{1}{\rho_{1}(\varepsilon)}\frac{d\rho_{1}(\varepsilon)}{d\varepsilon}\right)\right|_{\varepsilon=E^{*}_{1}}=\left.\left(\frac{1}{\rho_{2}(\varepsilon)}\frac{d\rho_{2}(\varepsilon)}{d\varepsilon}\right)\right|_{\varepsilon=E^{*}_{2}}=T^{-1}.
\end{eqnarray}
This expression can be used to define the temperature $T$ of the DNS. Although temperature is not explicitly present in the calculations, its value is useful for qualitatively assessing the rigidity of the DNS with respect to the excitation of various degrees of freedom.

In the calculation of $\rho_{1,2}$, the collective enhancement factors are not taken into account. The reason is that being constituents of DNS the collective motion of fragments are not independent of each other. Therefore, instead of collective enhancements of individual DNS fragments we take into account the collective excited states of DNS. Moreover, since the dynamics of DNS in fragments deformation and in relative distance are treated explicitly, the associated collective states are not incorporated to the DNS level densities. Therefore, the only collective states which contribute to the level densities are related to angular motion in DNS as discussed in Sec.~\ref{Sec.I}.

Denoting the energies of these collective states as $E_{x}^{*}$, ($x=0,1,2,...$), the DNS level density is obtained as a folding of intrinsic level density and the density of collective states as \cite{Rahmatinejad2020,Maino1990}
\begin{eqnarray}
\label{eq21}
\rho(E^{*},DNS)=\sum_x\rho_{int}(E^{*}-E_{x},DNS).
\end{eqnarray}
Note that if the excitation energy $E^{*}$ for a given DNS configuration  is small, then among the excitations of angular motions only a few contributes to $\rho(E^{*},DNS)$.

\subsection{Distribution of primary fission fragments \label{model2}}

The evolution of DNS distribution in time is described using the master equation. The space of deformations and masses of the fragments is discretized with grid sizes $\Delta \beta_{i}=0.05$, and $\Delta A_{i}=\Delta Z_{i}=1$ or $\Delta A_{i}=\Delta N_{i}=1$, respectively. Each of the obtained cells is characterized by the index
\begin{eqnarray}
n=(Z_{i},A_{i},\beta_{i}, i=1,2).
\end{eqnarray}
Defining the probability for the DNS to be in cell $n$ as $P(n)$, the time dependence of this probability is governed by the master equation \cite{Moretto1975}:
\begin{eqnarray}\label{M1}
 \frac{\textrm{d}}{\textrm{d}t}P(n,t)=\sum_{n'}\left[\Lambda(n'|n) P(n',t)- \Lambda(n|n')P(n,t)\right]-\Lambda_{decay}(n)P(n,t),
 \end{eqnarray}
with the  initial condition $P_{0}(n)=P(n,t=0)$ determined by the superposition of the initially formed DNS. In Eq.~\eqref{M1}, $\Lambda(n'|n)$ and $\Lambda(n|n')$ represent the rates of transitions from $n'$ to $n$ and vice versa. The DNS decay rate is denoted as $\Lambda_{decay}$.
Evolution is assumed to occur only between neighboring cells. The distribution of primary fission fragments $P_f(n[Z_1,A_1,\beta_1,Z_2,A_2,\beta_2])$ is obtained by solving Eq.(\ref{M1}) with Monte-Carlo technique.

The transition rates in Eq.~\eqref{M1} are expressed in terms of microscopic transition probabilities $\lambda_{nn'}$ and the level densities of the final states \cite{Moretto1975,Adamyan1994}:
\begin{eqnarray}\label{M2}
&\Lambda(n|n')=\lambda_{nn'}\rho(E^*_{n'},n'),   \nonumber\\
&\Lambda(n'|n)=\lambda_{n'n}\rho(E^*_{n},n), \nonumber\\
&\lambda_{n'n}=\lambda_{nn'}.
\end{eqnarray}
Microscopic transition probabilities are defined up to a constant parameter as \cite{Moretto1975}
\begin{eqnarray}\label{M3}
\lambda_{n'n}=\lambda_{nn'}=\lambda^{(j)}/\sqrt{\rho(E^*_{n'},n')\rho(E^*_{n},n)},\quad j=(A,b).
\end{eqnarray}
Here, $\lambda^{(A)}$ is used when $n$ and $n'$ differ by the fragment masses (nucleon transfer between the fragments), and $\lambda^{(b)}$ is used when $n$ and $n'$ differ by fragment deformations. In this study, $\lambda^{(A,b)}$ are treated as the parameters of the model and adjusted based on analysis of average neutron multiplicities in even-even $^{242-248}$Cm and $^{246-254}$Cf nuclei.

The decay probability in Eq.~\eqref{M1} is treated as
\begin{eqnarray}\label{M4}
\Lambda_{decay}(n)=\lambda^{d} \rho_{s.p.}(E_{n}^{*}-V_{B}^n,n),
\end{eqnarray}
where microscopic decay probability is
\begin{eqnarray}\label{M5}
\lambda^{d}=
\begin{cases}
    1/\rho(E_{n}^{*},n),&  \text{if } (E_{n}^{*}>V_{B}^n),\\
    0,              & \text{if } (E_{n}^{*}<V_{B}^n).
\end{cases}
\end{eqnarray}
In Eq.~\eqref{M4}, $\rho_{s.p.}(E_{n}^{*}-V_{B}^n,n)$ is the level density at the top of barrier in $R$ keeping the DNS. In Eq.~\eqref{M5}, tunneling through the potential barrier is disregarded. If excitation energy $E^{*}_n$ of DNS is less than the barrier height $V_{B}^n$, the decay channel is closed. If there is no barrier in the potential energy, then $\rho_{s.p.}(E_{n}^{*},n)$ is calculated at touching configuration $(V_{B}^n=0)$.

To choose the initial distribution $P_0(n,t=0)$ we select the DNS whose quadrupole moment $Q_{n}$
is within 10$\%$ range around the quadrupole moment $Q_{in}$ of an ellipsoid with an axes ratio $b_{in}$:
\begin{eqnarray} \label{eq10b}
|Q_{n}-Q_{in}|\leqslant 0.1 Q_{in}.
\end{eqnarray}
The value $b_{in}$ is treated as a parameter of the model.

Its choice is governed by two conditions. First, the value of $b_{in}$ must correspond to the elongations of the fissioning nucleus that exceed the position of the fission barrier. Secondly, at this point it should be possible to present the fissile nucleus as a superposition of binary systems (DNS). As follows from the observation that hyperdeformed states of nuclei ($b=3:1$) can be treated as DNS \cite{Shneidman2000,Nazarewicz1992,Cwiok1994,Aberg1994}, both these conditions can be fulfilled  for  $b_{in} > 3.0$. It is interesting that our calculations also show that the DNS which can potentially contribute to the initial state ($E^*_n>0$) starts to emerge in this region.
The chosen DNS are then distributed proportionally to their level densities:
\begin{eqnarray} \label{eq10c}
\frac{P_{0}(n)}{P_{0}(n')}=\frac{\rho(E^*_{n},n)}{\rho(E^*_{n'},n')}.
\end{eqnarray}

\subsection{Angular motions of fission fragments at scission \label{Sec.I}}

 In the model presented, it is assumed that fragments acquire angular momentum at scission due to the collective angular motion in DNS. Moreover, as discussed above, these states contribute to the collective enhancement of DNS level densities (see Eq.~\eqref{eq21}). The detailed quantum-mechanical description of this collective mode is presented in Ref.~\cite{Shneidman2024} to describe an absence of correlation between average angular momenta of fission fragments and the sawtooth behavior of average angular momentum versus mass of fragments.
Here, we present a brief description of the main aspects of the model.

\begin{figure}
    \centering  \includegraphics[width=0.7\linewidth]{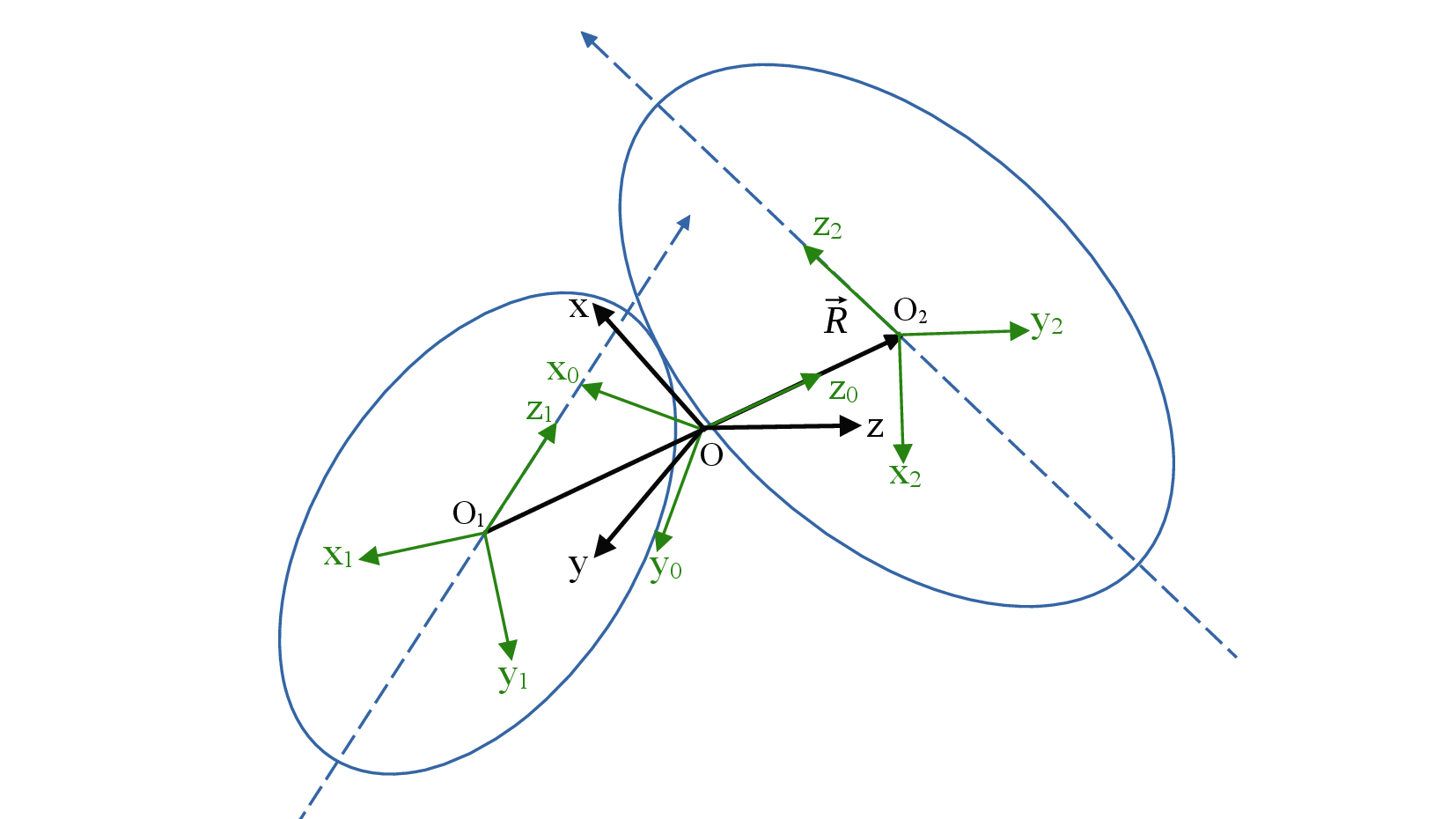}
    \caption{Schematic presentation of DNS or scission configuration of fissioning nucleus. The laboratory coordinate system $Oxyz$, as well as molecular $Ox_0y_0z_0$ and intrinsic coordinate systems $O_{i}x_iy_iz_i$, ($i=1,2$) of the fragments are marked. Center of laboratory and molecular coordinate systems $O$ coincide with the center of mass of DNS. Centers of intrinsic systems $O_{1,2}$ coincide with the center of mass of the corresponding fragments. The vector of distance between the centers of mass of the fragments is denoted as ${\bf R}$.}
    \label{Schemartic}
\end{figure}

The schematic representation of DNS ($Z_{1,2},A_{1,2},\beta_{1,2}$) is shown in Fig.~\ref{Schemartic}. The center of laboratory system  $Oxyz$ is placed in the center of mass of DNS. The vector of relative distance between the centers of masses of the fragments is denoted as ${\bf R}=(R,\theta_{0},\phi_{0})$, where $\Omega_0=(\theta_{0},\phi_0)$ are the angles determining the orientation of ${\bf R}$ with respect to the laboratory system.  The body-fixed (molecular) coordinate system $Ox_{0}y_{0}z_{0}$ is taken in such a way that the direction of  $Oz_{0}$ axis  coincides with vector ${\bf R}$. The centers of the intrinsic coordinate system of the $i$-th fragment $O_i$ are  placed in the centers of mass  of the corresponding fragment. The orientation of intrinsic coordinate systems $O_{i}x_{i}y_{i}z_{i}$ of $i$-th fragment with respect to the laboratory system is defined by the angles  $\Omega_{i}=(\phi_{i},\theta_{i},0)$. Since the fragments are assumed to be axially-symmetric, only two Euler angles are required.
The orientations of the fragments and of ${\bf R}$ with respect to the laboratory system  are not restricted. We assume that before decay,  fragments at any orientation stay in touching configuration defined by $R_{m}(\Omega_{0},\Omega_{1},\Omega_{2})$.
Although the distance $R_m$ between the fragments for a given orientation does not depend on the rotation of the system as a whole, it still depends on angles $(\theta_{0},\phi_{0})$, because all angles are defined with respect to the laboratory frame rather than to the body-fixed frame.

The  collective  motion in DNS is related to the rotation of the fragments and to the relative rotation of DNS as a whole with respect to the laboratory system. The kinetic energy is then written as
\begin{eqnarray}\label{eq1}
    T=\frac{\hbar^2\hat{I}_{0}^2}{2\mu R^2}+\frac{\hbar^2\hat{I}_{1}^2}{2\Im_{1}}+\frac{\hbar^2\hat{I}_{2}^2}{2\Im_{2}},
\end{eqnarray}
where angular momentum operators are given as
\begin{eqnarray} \label{eq2}
\hat{I}_{i}^2=-\left(\frac{1}{\sin\theta_{i}}\frac{\partial}{\partial\theta_{i}}\sin \theta_{i}\frac{\partial}{\partial\theta_{i}}+\frac{1}{\sin^2\theta_{i}}\frac{\partial^2}{\partial\phi_{i}^2}\right),\quad (i=0,1,2).
\end{eqnarray}
In Eq.~\eqref{eq1}, $\Im_{1,2}$, and $\Im_{R}=\mu R^2$  are the moments of inertia of the fragments and  DNS as a whole, respectively,  $\mu=m_{0}A_{1}A_{2}/A$ is the reduced mass, and  $m_{0}$ is the nucleon mass. The moments of inertia of the  fragments are taken in the rigid body limit as
\begin{eqnarray}
    \label{momin}
&&\Im_{i}=\int d{\bf \tau}(x^2_{i}+z^2_{i})\rho_i ({\bf r})=  \int d{\bf \tau}(y^2_{i}+z^2_{i})\rho_i ({\bf r}) \nonumber \\
&&\approx \frac{2}{5}m_0 A_i R_i^2 \left (1+\sqrt{\frac{5}{16\pi}}\beta_i+\frac{20}{7\pi}\beta_i^2+\frac{5\sqrt{5}}{8\pi^{3/2}}\beta_i^3+... \right ).
\end{eqnarray}
The integrations in Eq.~\eqref{momin} are performed in the intrinsic coordinate system $O_ix_iy_iz_i$ of the corresponding fragments. The second line of Eq.~\eqref{momin} is obtained under the assumption that the nuclear densities $\rho_i$ ($i=1,2$)  of the fragments in their respective intrinsic coordinate systems are  constant   in the volume defined by the   surface $R_i(\theta,\phi)=R_{0i}(1+\beta_{i} Y_{20}(\theta,\phi)).$ Here, $R_{0i}=r_0 A_i^{1/3}$ with $r_0=1.2$ fm. The assumption of the rigid body moment of inertia stems from the fact that even at low excitation energies, the nucleus transits from a superfluid to a
normal phase. This transition is known to result in an increase of the moment of inertia towards the rigid body limit. Thus, the use of rigid body moments of inertia is a realistic approximation.

The potential energy is calculated as a sum of nuclear $V_{N}$, and Coulomb interactions $V_{C}$:
\begin{eqnarray} \label{eq3}
V_{int}(\Omega_i)=V_{C}(R_{m}({\Omega}_{i}),\Omega_{i})+V_{N}(R_{m}({\Omega}_{i}),\Omega_{i}), \qquad (i=0,1,2)
\end{eqnarray}
at touching configuration $R=R_m(\Omega_0,\Omega_1,\Omega_2)$.
For a given distance $R$ between the fragments and the orientation defined by angles $\Omega_{i=0,1,2}$, the Coulomb interaction is calculated as a multipole expansion
\begin{eqnarray} \label{eq4}
V_{C}(R,\Omega_{i})&=&\sum_{\lambda_1\lambda_2}V_{\lambda_1,\lambda_2}(\beta_1,\beta_2)\left [ \left[Y_{\lambda_1}({\Omega}_1)\times Y_{\lambda_2}({\Omega}_2)\right]_{(\lambda_1+\lambda_2)} \times Y_{\lambda_1+\lambda_2}({\Omega}_0) \right ]_{(0,0)}, \nonumber \\
V_{\lambda_1,\lambda_2}(\beta_{1},\beta_{2})&=&(-1)^{\lambda_2}\sqrt{\frac{(4\pi)^3(2\lambda_{1}+2\lambda_{2})!}{(2\lambda_{1}+1)!(2\lambda_{2}+1)!}} \frac{Q_{\lambda_1}^{(1)}(\beta_{1})Q_{\lambda_2}^{(2)}(\beta_{2})}{R^{\lambda_1+\lambda_2+1}},
\end{eqnarray}
where multipole moments of the fragments are
\begin{eqnarray} \label{eq5}
Q_{\lambda}^{(i)}(\beta_{i})=\sqrt{\frac{4\pi}{2\lambda+1}}\int\rho_{i}(\textbf{r}_{i})r_{i}^{\lambda}Y_{\lambda 0}(\theta,\phi)d\textbf{r}_{i}.
\end{eqnarray}

The nuclear part of interaction is calculated using the double-folding procedure \cite{Adamian1996} with density-dependent effective nucleon–nucleon interaction \cite{Migdal1967}. The nuclear densities $\rho_i$ are approximated by Fermi distributions with the radius parameter $r_0$=1.15 fm. The diffuseness parameters are taken as $a=0.56 \sqrt{B^{(0)}_n/B_n}$, where $B^{(0)}_n$ and $B_n$ are the neutron binding energies of the nucleus under study and the heaviest isotope of the same element, respectively. Details of the calculations are presented in  Refs.~\cite{Shneidman2015,Adamian1996}.

Calculating potential energy at touching configuration for all orientations, we expand it in a basis of tripolar spherical functions
\begin{eqnarray} \label{eq6}
V_{int}(\Omega_{i=0,1,2})=\sum_{\lambda_{0},\lambda_{1},\lambda_{2}} C_{\lambda_{0},\lambda_{1},\lambda_{2}}(\beta_1,\beta_2)\left [ \left[Y_{\lambda_1}({\Omega}_1)\times Y_{\lambda_2}({\Omega}_2)\right]_{\lambda_0} \times Y_{\lambda_0}({\Omega}_0) \right ]_{(0,0)},
\end{eqnarray}
where the stiffness parameters $C_{\lambda_{0},\lambda_{1},\lambda_{2}}$ are obtained by fitting the calculated potential energy.
The tripolar spherical functions are defined as \cite{Varshalovich1988}
\begin{eqnarray}\label{eq7}
   && \left [ \left[Y_{\lambda_1}({\Omega}_1)\times Y_{\lambda_2}({\Omega}_2)\right]_{\lambda_{12}} \times Y_{\lambda_0}({\Omega}_0) \right ]_{(\lambda \mu)}\equiv \left [[ \lambda_{1}\times \lambda_{2}]_{\lambda_{12}}\times \lambda_{0}\right ]_{(\lambda \mu)} \nonumber \\
   &&= \sum_{\mu_0,\mu_1,\mu_2,\mu_{12}} C_{\lambda_1 \mu_1, \lambda_2\mu_2}^{\lambda_{12}\mu_{12}}C_{\lambda_{12} \mu_{12}, \lambda_0\mu_0}^{\lambda\mu}Y_{\lambda_1 \mu_1}({\Omega}_1)Y_{\lambda_2 \mu_2}({\Omega}_2)Y_{\lambda_0 \mu_0}({\Omega}_0).
\end{eqnarray}

The Hamiltonian for the angular motion in DNS is then written as
\begin{eqnarray} \label{eq8}
&&H=\hat{T}+V_{int}(\Omega_{i=0,1,2})-V_{int}(\Omega_{i=0,1,2}=0), \label{eq8a}\\
&&\hat{T}=\frac{\hbar^2\hat{I}_{0}^2}{2\mu R^2}+\frac{\hbar^2\hat{I}_{1}^2}{2\Im_{1}}+\frac{\hbar^2\hat{I}_{2}^2}{2\Im_{2}},  \label{eq8b}\\
&&V_{int}(\Omega_{i=0,1,2})=\sum_{\lambda_{0},\lambda_{1},\lambda_{2}} C_{\lambda_{0},\lambda_{1},\lambda_{2}}\left [ \left[\lambda_1 \times \lambda_2\right]_{\lambda_0} \times \lambda_0 \right ]_{(0,0)}.
\label{eq8c}
\end{eqnarray}
In the Hamiltonian \eqref{eq8}, the energy of pole-to-pole touching configuration is extracted from the DNS potential energy. The energy of pole-to-pole configuration is taken into account in Eq.~\eqref{ebin}. Therefore, the Hamiltonian (\ref{eq8}) only takes into account the orientation dependence of DNS potential energy.

For a state with angular momentum $I$, the Hamiltonian \eqref{eq8a} is diagonalized using a set of tripolar spherical functions
$\left[ {i_{0}}\times [i_{1}\times i_{2}]_{i_{12}} \right]_{(I,M)}$. For  spontaneous fission of $^{252}$Cf, we are interested only in the states  $I^{\pi}=0^{+}$. Therefore, the basis set is reduced and the eigenfunctions of $0^{+}$ states of the angular motion are given as
\begin{eqnarray} \label{eq9}
\Psi_{x}=\sum_{i_{0}i_{1}i_{2}}a^{(x)}_{i_{0}i_{1}i_{2}}\left[i_{0}\times\left[i_{1}\times i_{2}\right]_{i_{0}}\right]_{(00)}, \quad (x=0,1,2),
\end{eqnarray}
where $a^{(x)}_{i_{0}i_{1}i_{2}}$ are the probability amplitudes obtained by the numerical diagonalization of the Hamiltonian \eqref{eq8a}.
The expression
\begin{eqnarray} \label{eq10}
\mathcal{P}(x,i_{0}i_{1}i_{2})=|a^{(x)}_{i_{0}i_{1}i_{2}}|^2,
\end{eqnarray}
defines the probability  that the angular momentum of the first fragment, of the second fragment and of DNS as a whole  are $i_{1}$, $i_2$, and $i_0$, respectively.
Then the average angular momentum of the fragments and the binary system at the state $x$ is as follows
\begin{eqnarray} \label{eq11}
<i^{(x)}_{j}>=\left[\sum_{i_{0}i_{1}i_{2}} i_{j}(i_{j}+1)\mathcal{P}(x,i_{0}i_{1}i_{2})\right]^{1/2}, \quad (j=0,1,2).
\end{eqnarray}
For the  DNS characterized by excitation energy $E^*$, the angular momenta of the fragments are obtained by  averaging $<i^{(x)}_{j}>$
over different states $x$:
\begin{eqnarray} \label{eq12}
    <i_{j}>=\mathcal{N}\sum_{x}<i^{(x)}_{j}>\rho_{int}(E^{*}-E_{x},DNS),  \quad (j=0,1,2).
\end{eqnarray}
where
$\mathcal{N}^{-1}=\sum_x\rho_{int}(E^{*}-E_{x},DNS)$ is the normalization constant.


\subsection{Distribution of fission observables \label{model3}}
Solving Eq.~\eqref{M1}, we obtain the distribution of primary fission fragments $P_f(n[Z_1,A_1,\beta_1,Z_2,A_2,\beta_2])$.
The mass $Y(A)$ and charge $Y(Z)$ distributions are then calculated by summing over the probabilities of decay from the cells characterized by given mass $A$ or charge $Z$ values of one of the fragments, respectively:

\begin{eqnarray}\label{M7}
Y(A)=\sum_{n_j} P_{f}(n_j)\left[\delta_{A_1,A}+\delta_{A_2,A}\right],\nonumber \\
Y(Z)=\sum_{n_j} P_{f}(n_j)\left[\delta_{Z_1,Z}+\delta_{Z_2,Z}\right].
\end{eqnarray}

For a DNS characterized by index $n$, the TKE of primary fission fragments is defined as the interaction energy at the top of the barrier $R=R_{b}$:
\begin{eqnarray}\label{M7b}
 TKE_{n}=V_{C}(n[Z_{1,2},A_{1,2},\beta_{1,2},R_{b}])+V_{N}(n[Z_{1,2},A_{1,2},\beta_{1,2},R_{b}]).
\end{eqnarray}
Then the TKE distribution is calculated by summing over the probabilities $P_{f}(n)$ corresponding to the states with TKE within a given interval $2\Delta$:
\begin{eqnarray}\label{M7c}
Y(TKE)=\sum_{j} P_{f}(n_j)\Theta(|TKE_{n_j}-TKE|\leqslant \Delta),
\end{eqnarray}
where function $\Theta=1$ if the condition $|TKE_{n_j}-TKE|\leqslant \Delta$ is fulfilled and $\Theta=0$ otherwise.

To obtain the number of neutrons emitted by the fission fragments, we calculate their post-scission excitation energies. These energies originate from two sources: first, the excitation energies $E^{*}_{1,2}$ which fragments possess at scission (see Eq.~\eqref{M6b}); second, energies $E_{1,2}^{def}$ acquired by the fragments after scission due to the relaxation of their shapes back to the ground-state:
\begin{eqnarray} \label{DefEn}
E_{i}^{def}(n)&=&\left[U_{i}^{LD}(Z_i,A_i,\beta_{i})+\delta E_{i}^{sh}(Z_i,A_i,\beta_{i})\right]\nonumber \\ &-&\left[U_{i}^{LD}(Z_i,A_i,\beta_{i}^{gs})+\delta E_{i}^{sh}(Z_i,A_i,\beta_{i}^{gs})\right],\quad (i=1,2).
\end{eqnarray}

The probability $F_{1,2}(x,E_{1,2})$ that a fragment with a given excitation energy $E_{1,2}=E^*_{1,2}+E^{def}_{1,2}$ emits exactly $x$ neutrons  is determined through the  Monte-Carlo simulation applying probability densities derived from microscopic level densities and using experimental neutron separation energies (see Ref.~\cite{Rahmatinejad2023} for details).
The probability to emit $\nu$ neutrons in a fission event (neutron multiplicity distribution) is then given by the expression:
\begin{eqnarray} \label{eq7}
Y(\nu)=\sum_{j}\left[\sum_{x=1}^{\nu}F_{1}(x,E_{1}^{def}(n_j)+E_{1}^{*}(n_j))F_{2}(\nu-x,E_{2}^{def}(n_j)+E_{2}^{*}(n_j))\right] P_{f}(n_{j}).
\end{eqnarray}

Using the obtained yields $P_{f}(n)$, we calculate the angular momentum of a particular primary fission fragment as follows. For a given DNS indexed by $n$, the average angular momenta of first $\langle I_{1}(n) \rangle$ and second $\langle I_{2}(n) \rangle$ fragments and of the DNS as a whole $\langle I_{0}(n) \rangle $ are calculated using Eqs.~(\ref{eq11}), and (\ref{eq12}).

Then the average angular momentum of the given primary fission fragment $(Z,A)$ is calculated as:
\begin{eqnarray}\label{Eq8}
    \langle I_{Z,A} \rangle=N \sum_{n_j} P_{f}(n_j) \left[ \langle I_{1}(n_j) \rangle \delta_{Z_{1},Z}\delta_{A_{1},A}
     +\langle I_{2}(n_j) \rangle \delta_{Z_{2},Z} \delta_{A_{2},A} \right],
\end{eqnarray}
where $N^{-1}=\sum_{n_j} P_{f}(n_j) \left[\delta_{Z_{1},Z}\delta_{(A_{1},A}+\delta_{Z_{2},Z} \delta_{A_{2},A}\right]$ is the normalization constant.

If any additional restrictions on the choice of primary fragments, such as, for example, fixed deformation or TKE are imposed, they can be incorporated through additional constraints on the choice of $n$ in Eq.~\eqref{Eq8}. In order to link pre-scission and post-scission distributions one should additionally take into account the change of the fragment mass due to the neutron emission and the angular momentum carried away by neutrons. Following the discussion in~\cite{Giha2025}, we assume that each neutron carries away $\hbar/2$ units of angular momentum.

\section{Results \label{result}}

The main goal of this study is to investigate recently observed  dependence of angular momentum of $^{144}$Ba produced in spontaneous fission of  $^{252}$Cf on the TKE \cite{Giha2025}. For consistency, other fission observables are calculated as well.

\subsection{Mass, TKE, and neutron multiplicity distributions \label{result1}}

\begin{figure}
    \centering
    \includegraphics[width=0.48\linewidth]{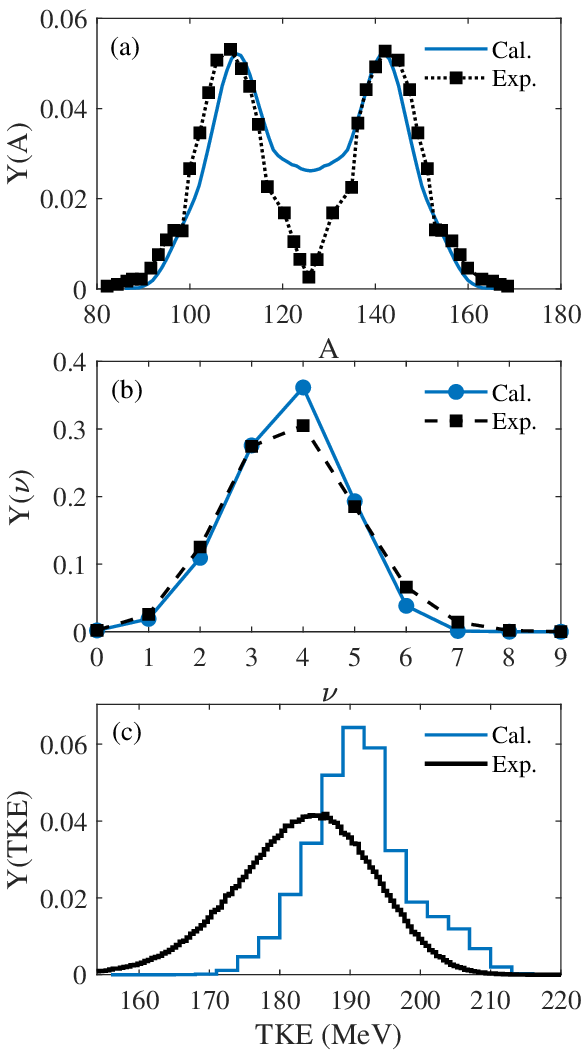}
    \caption{Calculated (solid lines) and experimental (dashed lines) mass (a), neutron multiplicity (b), and total kinetic energy (c) distributions for spontaneous fission of $^{252}$Cf. The experimental data for  mass is taken from \cite{David1976}, for neutron multiplicity  from \cite{Norman1986}, and  for TKE from \cite{Giha2025}.}
    \label{Pns}
\end{figure}

We first calculate mass, TKE and neutron multiplicity distributions. The results, together with the available experimental data, are shown in Figs.~\ref{Pns}(a-c). The average $\langle TKE \rangle$, the width $\langle A_{H} \rangle$ and light $\langle A_{L} \rangle $ fragments, and average neutron multiplicity $\langle \nu \rangle$ are listed in Table~\ref{Tab1}. In these calculations we use the following set of parameters: $\lambda^{A}=0.18 \text{s}^{-1}, \lambda^{b}=0.2 \text{s}^{-1}, b_{0}=3.55$.
Unlike in the model presented in Ref.~\cite{Rahmatinejad2024_2}, here the steps in mass and charge are taken as $\Delta A_{i}=\Delta Z_{i}=1$ or $\Delta A_{i}=\Delta N_{i}=1$. This was done in order to take into account the contribution from DNS with odd-mass fragments.
The results are in satisfactory agreement with the experimental data.

 \begin{table}
    \caption{The calculated and experimental average $\langle TKE \rangle$, the width of TKE distribution $\sigma_{TKE}$, average masses of heavy $\langle A_{H} \rangle$ and light $\langle A_{L} \rangle$ fragments together with the average neutron multiplicity $\langle \nu \rangle $.  \label{Tab1}}
             \centering
    \begin{tabular}{|c|c|l|}
        \hline
        Quantities & Exp. & Cal. \\
        \hline
        \multirow{2}{*}{$\langle TKE\rangle$ (MeV)} & $189\pm1$ \cite{TerAkopian1997} & 189.93 \\
        & 186.8 \cite{David1976} &  \\
        \hline
        \multirow{2}{*}{$\sigma_{TKE}$ (MeV)} & 9.43 \cite{TerAkopian1997} & 7.36 \\
        & 11.17 \cite{David1976} &  \\
        \hline
        \multirow{2}{*}{$\langle A_{H} \rangle $} & $145.7\pm0.1$ \cite{TerAkopian1997} & 140.41 \\
        & 143.4 \cite{David1976} &  \\
        \hline
        \multirow{2}{*}{$ \langle A_{L} \rangle $} & $106.3\pm0.1$ \cite{TerAkopian1997} & 111.59 \\
        & 108.6 \cite{David1976} &  \\
        \hline
        \multirow{2}{*}{$\langle \nu \rangle $} & $3.757\pm0.01$ \cite{Norman1986} & 3.713 \\
        & $3.756\pm0.031$ \cite{Vorobyev2005} &  \\
        \hline
    \end{tabular}
\end{table}

 For mass distribution,  symmetric fission is  more pronounced comparing to the experimental data. This is also reflected in that the average heavy fragment mass is overestimated by 3-5 units compared to the experimental data (see Table.~\ref{Tab1}). However, in the region of Ba+Mo mass split (which is of main importance for the present study) our calculations are in a good agreement with the experimental mass distribution.

 Since, the average neutron multiplicities are used to define the parameters of the model, the neutron multiplicity distribution is reproduced with a good accuracy (see Fig.~\ref{Pns} (b)). The width of this distribution is also well reproduced. Previously, the model was successfully applied to the description of recent results on neutron multiplicities in transfermium nuclei \cite{244Fm-JINR,Isaev2023,Mukhin2024-2}.

The TKE distribution $Y(\textrm{TKE})$ is presented in Fig.~\ref{Pns} (c) together with experimental data from Ref.~\cite{Giha2025}. It is seen that the maximum of calculated distribution is shifted about 5 MeV towards larger TKE and the distribution is narrower to the direction of small TKE.

 It is interesting that our $\langle TKE \rangle$ aligns with the average TKE which is given for the first mode in Ref.~\cite{TerAkopian1997} (see Table.~\ref{Tab1}). In this experiment, two modes were detected for the  TKE  and neutron emission. Within the configuration space considered, we do not observe the second mode which corresponds to a  large number of neutrons. Since our aim is to analyze the angular momentum versus TKE in the range of 170-200 MeV, an absence of second mode does not affect our analysis.

\subsection{Scission DNS configurations leading to $^{144}$Ba fission fragment \label{result2}}

In order to investigate angular momentum of $^{144}$Ba as a function of TKE, we should first determine which DNS configurations lead to the post-scission $^{144}$Ba fragment. The proton emission is strongly hindered, therefore, we restrict the consideration by Ba+Mo scission configurations only.
For example, post-scission $^{144}$Ba can be obtained from the DNS: $^{144}$Ba+$^{108}$Mo at scission under constraint that no neutrons were emitted from Ba after decay.

The post-scission $^{144}$Ba is also produced from the following scission configurations:
\begin{eqnarray}\label{Eq36}
&&^{144}\text{Ba}+^{108}\text{Mo}:\quad n_{\text{Ba}}<1,\nonumber \\
&&^{145}\text{Ba}+^{107}\text{Mo}:\quad 1 \leqslant n_{\text{Ba}}<2,\nonumber \\
&&^{146}\text{Ba}+^{106}\text{Mo}:\quad 2 \leqslant n_{\text{Ba}}<3,\nonumber \\
&&^{147}\text{Ba}+^{105}\text{Mo}:\quad 3 \leqslant n_{\text{Ba}}<4,
\end{eqnarray}
where $n_{\text{Ba}}$ is the average number of neutrons emitted from Ba fragment.
Other scission configurations leading to post-scission  $^{144}$Ba are populated with quite small probabilities and are not taken into account.

\begin{figure}[t]
    \centering  \includegraphics[width=0.48\linewidth]
    {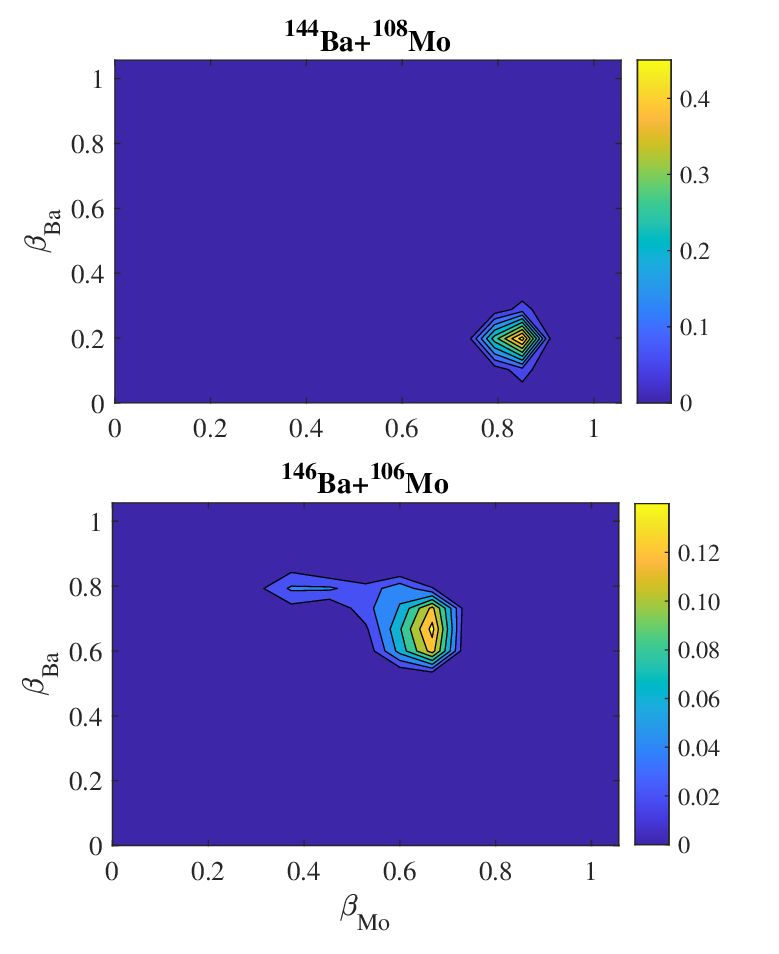}
     \centering  \includegraphics[width=0.48\linewidth]
     {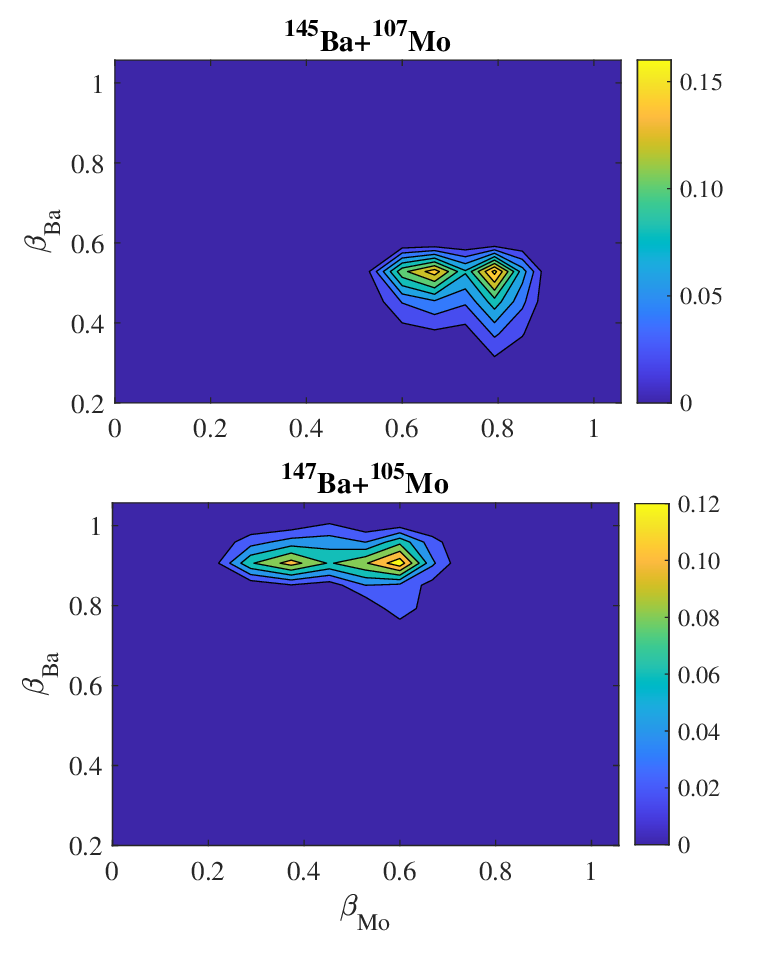}
    \caption{The decay probabilities of various scission configurations leading to post-scission fragment $^{144}$Ba as a function of the deformations $\beta_{\textrm{Mo}}$ and $\beta_{\textrm{Ba}}$ of the Mo and Ba fragments, respectively.}
    \label{147Ba145Bab1b2}
\end{figure}

The probabilities that $^{252}$Cf  decays from the scission configurations \eqref{Eq36} are presented in Fig.~\ref{147Ba145Bab1b2} as a function of the deformations $\beta_{\textrm{Mo}}$ and $\beta_{\textrm{Ba}}$ of the Mo and Ba fragments, respectively. Since the amount of neutrons emitted from Ba is constrained, the decay occurs from the region limited in $\beta_{\text{Ba}}$.
Indeed, following the discussion regarding Eq.~\eqref{eq7}, the excitation energy required for neutron emission comes from the excitation energy at scission and deformation energy. These energies increase with deformation of fragments. Therefore, restriction on the number of neutrons emitted imposes the restriction on the deformation of Ba at scission configuration.

The number of neutrons emitted from Mo is not constrained. That is why the decay can occur in a wider range of $\beta_{\text{Mo}}$.

\begin{figure}    \centering  \includegraphics[width=0.9\linewidth]{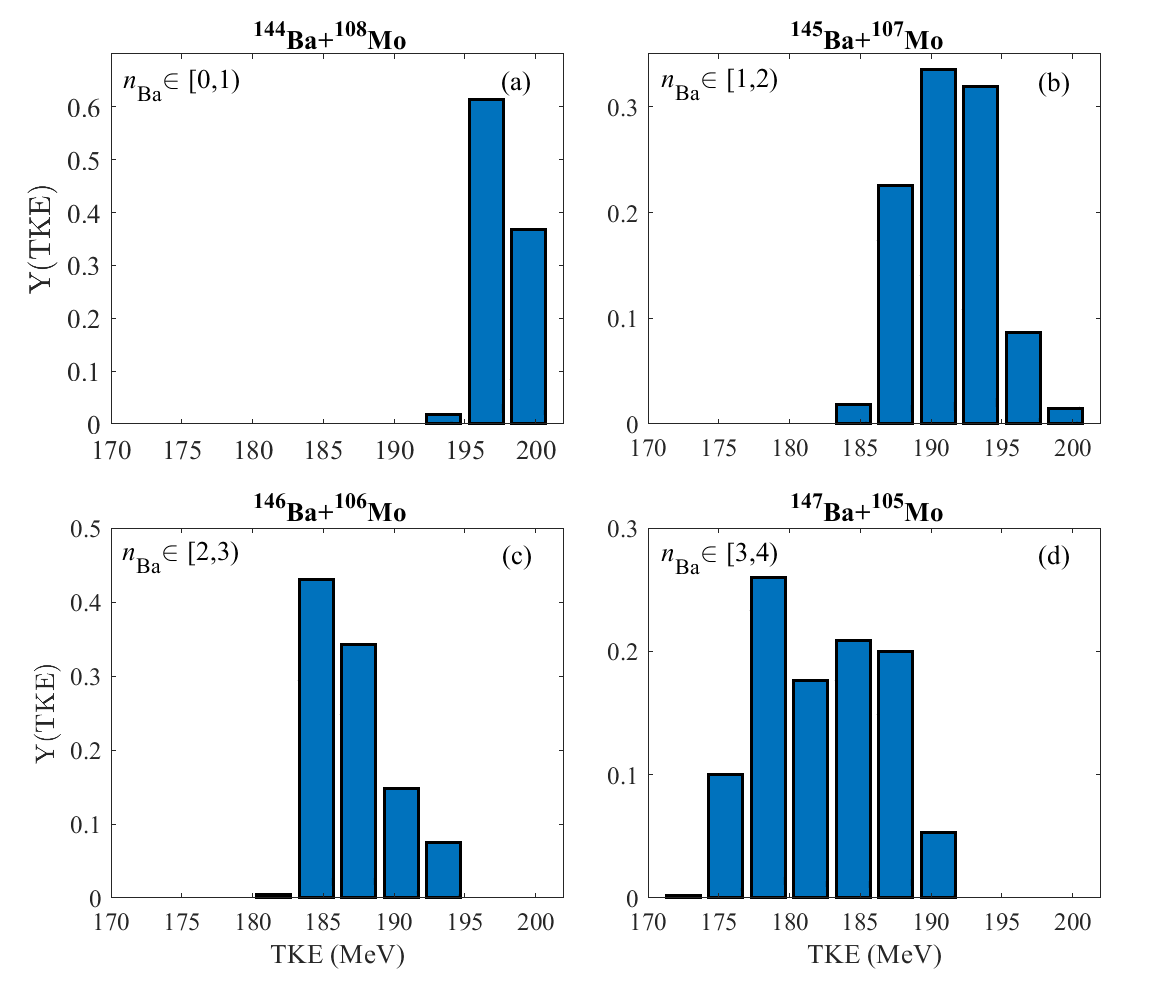}
    \caption{The calculated total kinetic energy distributions for the decay from various scission configurations leading to post-scission fragment $^{144}$Ba. Each TKE distribution presented here is normalized to unity for better visibility. }
    \label{BaTKEbar-4plot}
\end{figure}

The decay of DNS with given deformations leads to fission event with TKE which is calculated with Eq.~\eqref{M7b}.
The TKE distributions $Y(TKE)$ for the DNS configurations presented in Eq.~\eqref{Eq36} are shown in Fig.~\ref{BaTKEbar-4plot}.
The TKE range is divided into bins with a width of 3 MeV.
The $Y(TKE)$ is calculated by summing up the decay probabilities leading to the  TKE values within the range $\pm 1.5$ MeV. Each TKE distribution presented in Fig.~\ref{BaTKEbar-4plot} is normalized to unity for better visibility. As seen from the figure, each DNS has a TKE distribution which is dominant in a certain energy interval. In some cases these intervals overlap with each other. Then, several DNS contribute to the distribution of fission observables at a given TKE simultaneously. The contribution of these systems is determined by their relative decay probabilities.

\subsection{Description of angular vibrations in scission DNS configurations \label{result3}}

For a given TKE bin, the deformations of the DNS fragments change only slightly.  Therefore, in the calculation of angular momenta of fission fragments, the deformations of the DNS constituents are averaged inside each bin. The average deformations for each TKE bin together with the corresponding TKE values, decay probabilities and temperatures at  scission point for the chosen DNS are listed in Table.~\ref{tab2}. For these systems  we calculate the potential energy for angular vibrations using Eqs.~(\ref{eq3}), and (\ref{eq6}). Diagonalizing the Hamiltonian in Eqs.~(\ref{eq8a}-\ref{eq8c}), we obtain the spectra of collective $0^+$-states corresponding to angular motions in DNS. For each state $x$, we calculate the average angular momenta $\langle I^{(x)}_{\textrm{Mo}}\rangle $, $\langle I^{(x)}_{\textrm{Ba}} \rangle$, and $\langle I^{(x)}_{0} \rangle$. Finally, we average over different collective states using Eq.~\eqref{eq12}. The obtained values of $ \langle I_{\textrm{Mo}}\rangle$, $\langle I_{\textrm{Ba}}\rangle$, and $\langle I_{0} \rangle$ are also presented in Table.~\ref{tab2}.

\begin{figure}
  \centering
   \fbox{
\includegraphics[width=0.4\linewidth]{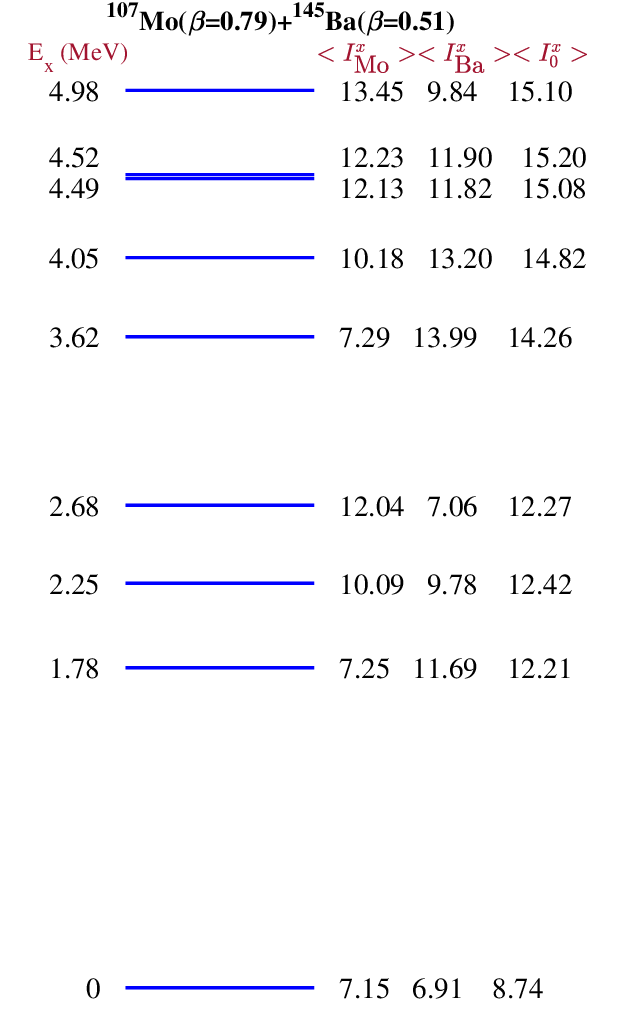}}
    \caption{The spectrum of angular motion in DNS: $^{107}$Mo$(\beta_{\textrm{Mo}}=0.79)+^{145}$Ba$(\beta_{\textrm{Ba}}=0.51)$ is shown. The excited states are marked by the lines, with average angular momenta $\langle I_{\textrm{Mo}}\rangle$, $\langle I_{\textrm{Ba}}\rangle$ and $\langle I_{0}\rangle$ listed to the right and excitation energies $E_{x}$ to the left of each level.}
    \label{Spec107Mo145Ba}
\end{figure}

As an example, the spectrum of DNS: $^{107}$Mo$(\beta_{\textrm{Mo}}=0.79)+^{145}$Ba$(\beta_{\textrm{Ba}}=0.51)$ is displayed in Fig.~\ref{Spec107Mo145Ba}. To the right side of each excited level $x$, average angular momenta $\langle I^{(x)}_{\textrm{Mo}}\rangle$, $\langle I^{(x)}_{\textrm{Ba}}\rangle$ and $\langle I^{(x)}_{0}\rangle$ are given.
To the left side of each level the excitation energies $E_{x}$ are indicated. As seen,  the average angular momenta grows  with  increasing $E_{x}$. However, the first excited state lies at 1.78 MeV. If we estimate the temperature $T$ of DNS using the Eq.~\eqref{M6b}, we obtain $T=0.44$ MeV which is sufficiently smaller than $E_x$. Therefore, only the ground-state  contributes significantly  to the generation of angular momenta of  fragments. The same holds true for other  DNS studied (see  Table.~\ref{tab2} for comparison between $E_1$ and $T$).

\begin{figure}
    \centering  \includegraphics[width=0.6\linewidth]{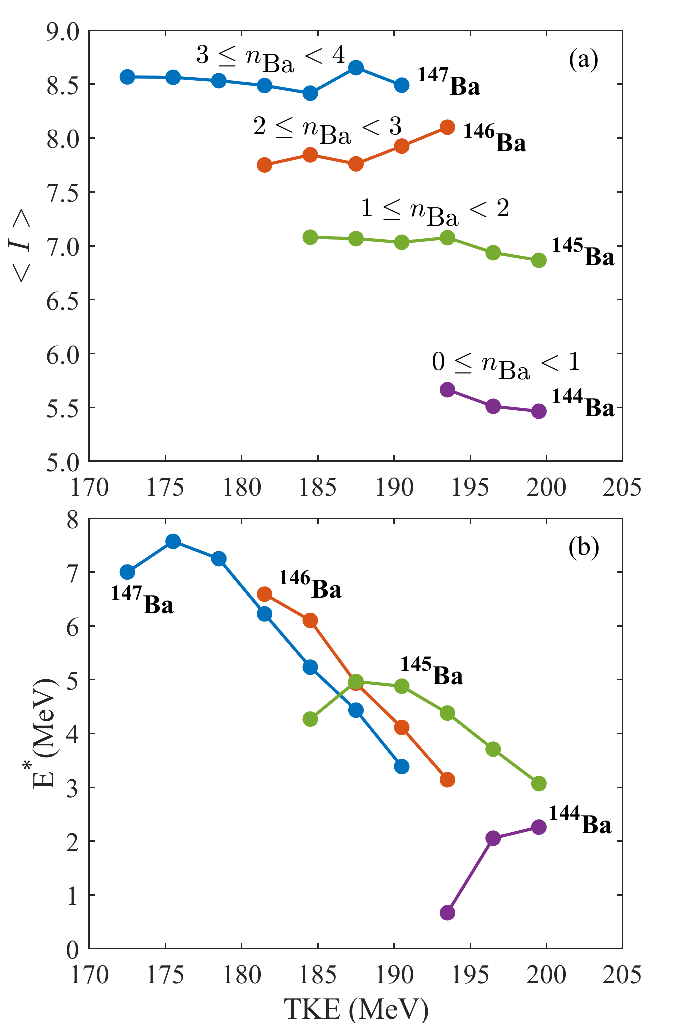}
    \caption{The calculated average angular momenta (panel a) and excitation energies (panel b) of Ba isotopes in scission configurations leading to $^{144}$Ba after $n_{\textrm{Ba}}$ neutron emissions, plotted against total kinetic energy TKE.}
    \label{TKE-I-total-raw}
\end{figure}

\subsection{Average angular momenta of primary fission fragments \label{result4}}

The calculated average angular momenta of Ba isotopes which lead to $^{144}$Ba after neutron emission are shown in Fig.~\ref{TKE-I-total-raw} (a).
As seen, the average angular momenta $ \langle I_{\textrm{Ba}} \rangle$ versus TKE are almost constant for each scission configuration.
As mentioned above, since the number of neutrons emitted from Ba isotopes is fixed, the deformation of Ba changes only slightly while variation in TKE is achieved due to change of deformation of the corresponding Mo isotopes. As shown in Ref.~\cite{Shneidman2024}, the average angular momenta of Mo and Ba pairs are uncorrelated and the average angular momenta of Ba isotopes stay almost constant within the available TKE range. Additionally, angular momentum increase due to excitation of higher states of angular motions is almost negligible because the excitation energies of the considered configurations are rather small (see Fig.~\ref{TKE-I-total-raw} (b)).

While for each scission configuration, the average angular momentum of the corresponding Ba isotope is practically constant, it differs from one scission configuration to another. Having maximum value ($\langle I_{\textrm{Ba}} \rangle \approx 8.65$) for the DNS: $^{147}$Ba$+^{105}$Mo, it decreases up to $\langle I_{\textrm{Ba}} \rangle \approx 5.46$ for DNS: $^{144}$Ba$+^{108}$Mo. The reason for this is clearly seen from  Fig.~\ref{147Ba145Bab1b2} where decay probabilities for various DNS configurations leading to post-scission $^{144}$Ba are displayed in the plane ($\beta_{\textrm{Ba}}$,$\beta_{\textrm{Mo}}$). The decay probability is localized in different regions of deformation of Ba fragment. For $^{144}$Ba pre-scision fragment, decay occurs mainly from $\beta_{\textrm{Ba}}\sim 0.2$, for $^{145}$Ba -- $\beta_\textrm{Ba}\sim 0.4$, for $^{146}$Ba -- $\beta_{\textrm{Ba}}\sim 0.7$, and for $^{147}$Ba -- $\beta_{\textrm{Ba}}\sim 0.92$.
As explained in details in Ref.~\cite{Shneidman2024}, the average angular momentum of a DNS fragment increases with its deformation.

\subsection{Average angular momenta of $^{144}$Ba and accompanying Mo fission fragments versus TKE \label{result5}}

If several DNS configurations contribute to the average angular momenta within a given TKE interval, we average them out using their corresponding decay probabilities presented in Table.~\ref{tab2}. Following experimental observations and discussion in Refs.~\cite{Giha2025}, neutrons are assumed to be emitted as $s$-waves. Therefore, each neutron is assumed to carry away $\hbar/2$ unit of angular momentum. Hence, to calculate average angular momentum of post-scission $^{144}$Ba, in addition to averaging we should take into account the different number of emitted neutrons from the  pre-scission Ba fragment:
\begin{eqnarray}
\langle I_A \rangle &=& \langle I({^A\textrm{Ba}})\rangle - (A-144)/2, \nonumber \\
    \langle I \rangle&=&\sum_{A=144,145,146,147} P_d(^{A}{\textrm{Ba}})\langle I_A \rangle /\sum_{A=144,145,146,147} P_d(^{A}{\textrm{Ba}}).
\end{eqnarray}

\begin{figure}
    \centering  \includegraphics[width=0.6\linewidth]{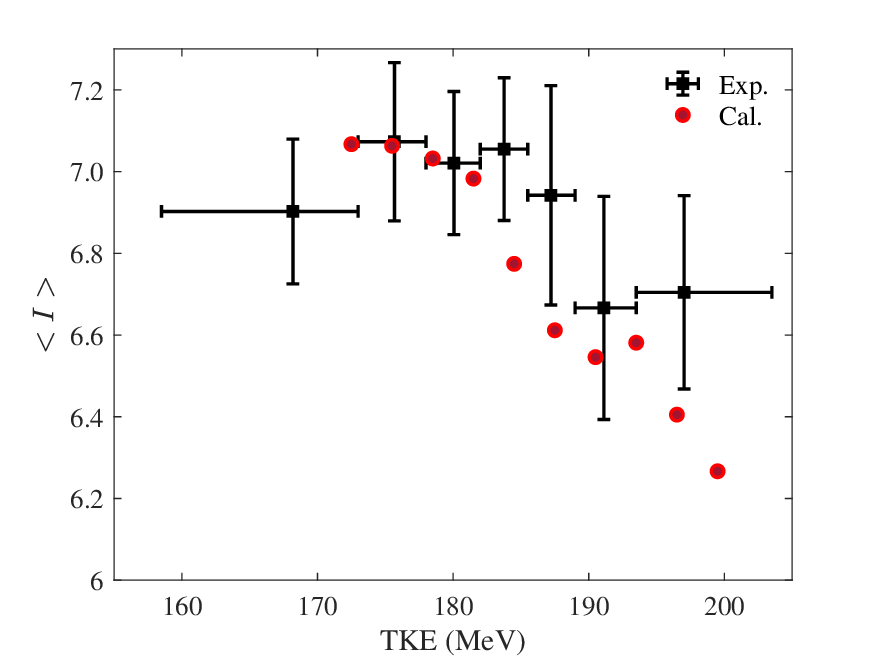}
    \caption{Average angular momentum $\langle I \rangle$ of post-scission $^{144}$Ba (red circles) as a function of total kinetic energy (TKE), compared with experimental data from Ref.~\cite{Giha2025} (black squares).}
    \label{TKE-I-total}
\end{figure}

The result of calculations for average angular momentum  $\langle I \rangle$ of post-scisson $^{144}$Ba versus TKE is presented in Fig.~\ref{TKE-I-total}, together with the experimental data  from Ref.~\cite{Giha2025}.
The absolute value of average angular momentum is in a good agreement with the experimental data and its trend is well reproduced by the calculations.
The change of angular momentum in the range of TKE from 173 MeV up to 200 MeV is limited to 0.8 $\hbar$. In the TKE range of 170-180 MeV the $ \langle I \rangle$ values stay constant. This is related to the fact that in this range of TKE only one scission configuration $^{147}$Ba$+^{105}$Mo  contributes. The contributions of the DNS $^{146}$Ba$+^{106}$Mo starting from TKE=181.5 MeV, and of the DNS $^{145}$Ba$+^{107}$Mo starting from TKE=184.5 MeV lead to gradual but slight decrease of angular momentum with increasing TKE. The local minimum is achieved at TKE=190.5 MeV where the contribution of the DNS $^{147}$Ba$+^{105}$Mo ends. Further increase of TKE leads to slight increase in average angular momentum because of the contrbution of the  DNS $^{146}$Ba$+^{106}$Mo. This up-bending at TKE $\approx$  190 MeV is also seen in the experimental data. Finally, when the contribution of DNS $^{146}$Ba$+^{106}$Mo comes to a halt at TKE=193.5 MeV, the decrease of $ \langle I \rangle$ continues. The decrease of angular momentum in the region of $\textrm{TKE}>193.5$ is related to the contribution of two DNS $^{145}$Ba$+^{107}$Mo and $^{144}$Ba$+^{108}$Mo. Note that the relative yield of the later DNS configuration is at least one order of magnitude smaller (see Table~\ref{tab2}) which possibly makes it hard to observe experimentally the events related to the decay of this system. If this DNS is excluded from our calculations the dependence of angular momentum on TKE gets even weaker.

To summarize, the model presented is capable to explain the weak dependence of average angular momentum on TKE for the post-scission isotope $^{144}$Ba. There are three  reasons for this weak dependence. Firstly, the change in TKE at a constraint of certain number of emitted neutrons from Ba isotope mainly occurs due to the variation of deformation of the corresponding Mo fragment. Secondly, the most of energy available for neutron emission is stored as deformation energy while the excitation energy at scission is rather small and the excitation of higher-lying states of collective angular motions is strongly hindered. Finally, the larger number of emitted neutrons leads to the  decrease of  angular momentum of post-scission fragment.

The observation that the change of TKE is mainly due to variation of deformation of Mo fragments gives an interesting possibility to check the validity of the model proposed.
If in coincidence with measurement of $^{144}$Ba the angular momentum of all Mo isotopes is measured, its trend versus TKE is expected to be different from that for $^{144}$Ba.

\begin{figure}
    \centering  \includegraphics[width=0.6\linewidth]{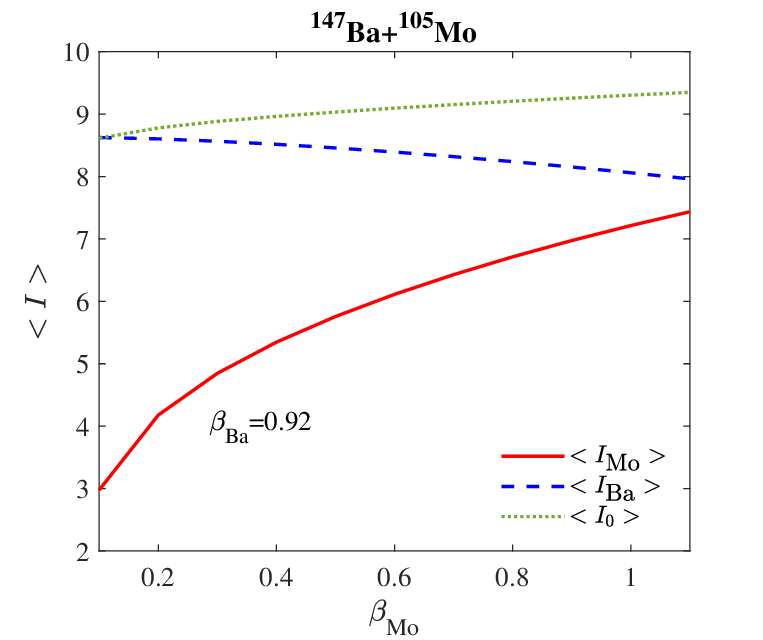}
    \caption{Calculated average angular momenta of pre-scission $^{105}$Mo ($\langle I_{\text{Mo}} \rangle$), $^{147}$Ba ($\langle I_{\text{Ba}} \rangle$), and the DNS as a whole ($\langle I_0 \rangle$) versus $\beta_{\text{Mo}}$. The deformation of $^{147}$Ba is fixed at $\beta_{\text{Ba}} = 0.92$, corresponding to the maximum decay probability for the DNS: $^{147}$Ba + $^{105}$Mo (see Fig.~\ref{147Ba145Bab1b2}).}
    \label{Mobetadependence}
\end{figure}

In Fig.~\ref{Mobetadependence}, we present the average angular momentum of pre-scssion $^{105}$Mo as a function of $\beta_{\textrm{Mo}}$.
The deformation of Ba is kept at  $\beta_{\textrm{Ba}}=0.92$ which corresponds to the maximum decay probability of the  DNS $^{147}$Ba$+^{105}$Mo (see Fig.\ref{147Ba145Bab1b2}). We see that average angular momentum of the pre-scssion $^{105}$Mo fragment exhibits a rapid increase at small deformations. This trend eventually ends up with an asymptotic at larger deformations.
After accounting for the angular momentum carried away by the emitted neutrons, which tend to increase at larger deformations, we anticipate a maximum average angular momentum, $\langle I \rangle$, for Mo at intermediate deformations.

\begin{figure}
    \centering  \includegraphics[width=0.6\linewidth]{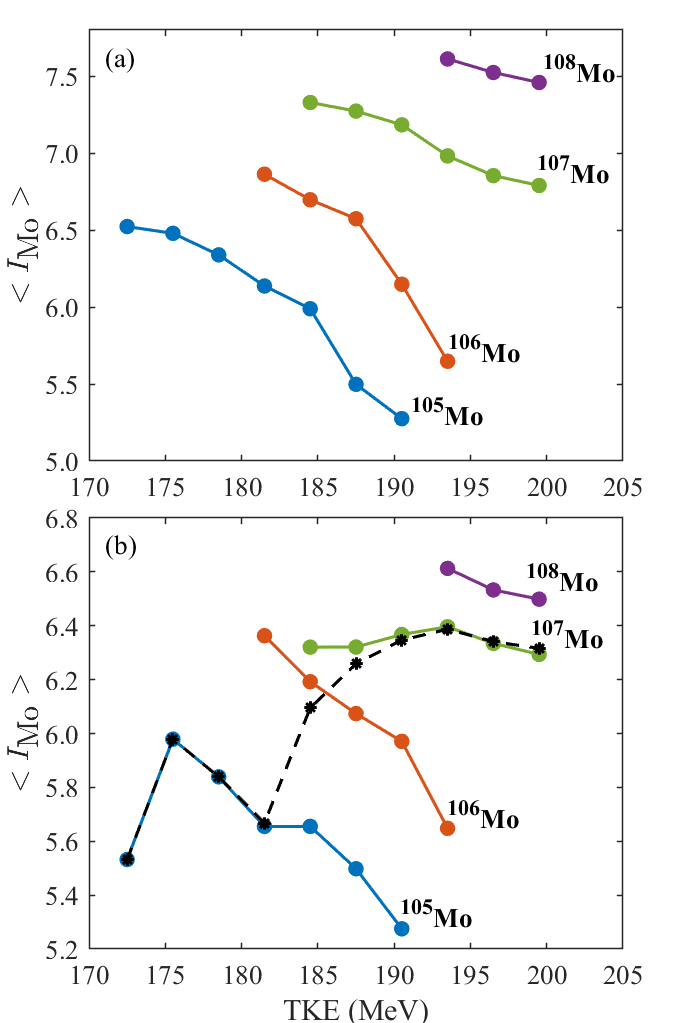}
    \caption{Calculated average angular momentum of pre-scission (panel a) and post-scission (panel b) Mo isotopes in the DNS configurations leading to $^{144}$Ba after neutron emission versus the total kinetic energy TKE. The dashed black line represents the summed contribution of various DNS configurations at scission, corresponding to the average angular momentum of Mo isotopes measured in coincidence with $^{144}$Ba versus TKE.}
    \label{TKE-I-total-raw-Mo}
\end{figure}

The angular momenta of pre-scission Mo isotopes are shown in Fig.~\ref{TKE-I-total-raw-Mo} (a). As seen, the angular momenta decrease gradually with increasing TKE (decreasing Mo deformation). After accounting for the angular momenta carried away by emitted neutrons the decline of angular momentum with TKE is generally reduced (see Fig.~\ref{TKE-I-total-raw-Mo} (b)). It is specially interesting in the case of $^{105}$Mo which exhibits a maximum at TKE$\approx 175$ MeV which appears due to the competition between increase of angular momentum and the number of emitted neutrons with deformation. Summing up the contribution of different DNS at scission we obtain the dashed black line which corresponds to average angular momentum of Mo isotopes versus TKE measured in coincidence with $^{144}$Ba.

It is seen that angular momentum of Mo varies in a wider range ($\approx 1\hbar$) than that of $^{144}$Ba. While the angular momentum of $^{144}$Ba stays almost constant and then slightly decreases with increasing TKE, the trend of Mo average angular momentum is opposite. Finally, in the region of low TKE, the angular momentum of Mo isotopes shows a local maximum at TKE $\approx175$ MeV.
Therefore, it seems interesting to investigate the average angular momentum of all Mo isotopes measured in coincidence with $^{144}$Ba.

\begin{table}[h]
\centering
\caption{The average deformations for each TKE bin together with the corresponding TKE values at the center of bins, decay probabilities and temperatures at the scission point for the chosen DNS.}
\label{tab2}
\begin{tabular}{|c|c|c|c|c|c|c|c|c|}\hline
 \multicolumn{9}{|c|}{$^{105}$Mo+$^{147}$Ba}\\
\hline
$\beta_\textrm{Mo}$ & $\beta_\textrm{Ba}$ & Yield & TKE (MeV) & T (MeV) & $\langle I_{\textrm{Mo}}\rangle \hbar$ & $\langle I_{\textrm{Ba}}\rangle \hbar$ & $\langle I_{\textrm{0}}\rangle \hbar$ & $E_1$ (MeV)\\
\hline
0.73 & 1.01 & 5.97$\times10^{-7}$& 172.50 & 0.53 & 6.52 & 8.57 & 9.37 & 2.14 \\
\hline
0.67 & 0.96 & 3.02$\times 10^{-5}$ & 175.50 & 0.56 & 6.48 & 8.56 & 9.39 & 2.10 \\
\hline
0.62 & 0.94 & 7.84$\times 10^{-5}$ & 178.50 & 0.54 & 6.34 & 8.53 & 9.32 & 2.05 \\
\hline
0.57 & 0.92 & 5.33$\times 10^{-5}$ & 181.50 & 0.50 & 6.14 & 8.49 & 9.21 & 1.98 \\
\hline
0.53 & 0.89 & 6.31$\times 10^{-5}$ & 184.50 & 0.45 & 5.99 & 8.42 & 9.11 & 1.98 \\
\hline
0.40 & 0.91 & 6.02$\times 10^{-5}$ & 187.50 & 0.43 & 5.50 & 8.65 & 9.04 & 1.67 \\
\hline
0.36 & 0.89 & 1.60$\times 10^{-5}$ & 190.50 & 0.37 & 5.28 & 8.49 & 8.95 & 1.59 \\
\hline
 \multicolumn{9}{|c|}{$^{106}$Mo+$^{146}$Ba}\\\hline
0.76 & 0.73 & 9.56$\times 10^{-7}$ & 181.50 & 0.44 & 6.86 & 7.75 & 9.02 & 2.14 \\ \hline
0.70 & 0.74 & 9.78$\times 10^{-5}$ & 184.50 & 0.44 & 6.70 & 7.84 & 9.02 & 2.20 \\ \hline
0.65 & 0.70 & 7.77$\times 10^{-5}$ & 187.50 & 0.39 & 6.57 & 7.76 & 8.92 & 2.19 \\ \hline
0.53 & 0.73 & 3.36$\times 10^{-5}$ & 190.50 & 0.36 & 6.15 & 7.93 & 8.85 & 2.09 \\ \hline
0.41 & 0.77 & 1.71$\times 10^{-5}$ & 193.50 & 0.33 & 5.65 & 8.10 & 8.80 & 1.81 \\
\hline
 \multicolumn{9}{|c|}{$^{107}$Mo+$^{145}$Ba}\\ \hline
0.87 & 0.55 & 8.08$\times 10^{-5}$ & 184.50 & 0.40 & 7.33 & 7.08 & 8.92 & 1.79 \\ \hline
0.83 & 0.53 & 9.66$\times 10^{-4}$ & 187.50 & 0.43 & 7.27 & 7.07 & 8.90 & 1.78 \\ \hline
0.79 & 0.51 & 1.44$\times 10^{-3}$ & 190.50 & 0.43 & 7.18 & 7.03 & 8.84 & 1.78 \\ \hline
0.71 & 0.50 & 1.37$\times 10^{-3}$ & 193.50 & 0.40 & 6.98 & 7.08 & 8.77 & 1.82 \\ \hline
0.67 & 0.47 & 3.71$\times 10^{-4}$ & 196.50 & 0.37 & 6.85 & 6.94 & 8.63 & 1.81 \\ \hline
0.63 & 0.44 & 6.27$\times 10^{-5}$ & 199.50 & 0.34 & 6.79 & 6.87 & 8.57 & 1.77 \\
\hline
 \multicolumn{9}{|c|}{$^{108}$Mo+$^{144}$Ba}\\ \hline
0.91 & 0.29 & 3.86$\times 10^{-7}$ & 193.50 & 0.14 & 7.61 & 5.66 & 8.50 & 1.27 \\ \hline
0.86 & 0.24 & 1.28$\times 10^{-5}$ & 196.50 & 0.28 & 7.52 & 5.51 & 8.40 & 1.19 \\ \hline
0.83 & 0.23 & 7.68$\times 10^{-6}$ & 199.50 & 0.28 & 7.46 & 5.46 & 8.34 & 1.18 \\
\hline
\end{tabular}
\end{table}

\section{Conclusions \label{conclusion}}
We propose a model to describe the evolution of a fissioning nucleus after tunneling through the fission barrier as a random walk among various scission configurations.
The nucleus is represented as a superposition of various DNS. The competition between DNS decay and its evolution in mass/charge asymmetry and fragments deformations leads to the formation of primary fission fragment distribution. The relative probabilities of initially formed DNS and the transition rates between them are determined by their level densities. The collective states associated with angular motion in the DNS are treated quantum-mechanically, as in Ref.~\cite{Shneidman2024}, and are included in the calculations. The model is applied to determine key fission observables in the SF of $^{252}$Cf. The calculated mass, TKE, and neutron multiplicity distributions agree well with experimental data. Further, for different values of TKE, we identify the DNS configurations yielding the post-scission fragment $^{144}$Ba after neutron emission. For these configurations, the average angular momenta of the DNS fragments are computed. By summing the contributions of the selected DNS and assuming that neutrons are emitted as $s$-waves, we determine the dependence of the post-scission $^{144}$Ba angular momentum on TKE. The results well describe the recent experimental data from Ref.~\cite{Giha2025}, confirming the weak correlation between fission fragment angular momenta and TKE. Additionally, we present the average angular momentum dependence for Mo fission fragments measured in coincidence with $^{144}$Ba.

\section*{Acknowledgments}
This work was supported by a grant \textnumero 25-42-00018 from the Russian Science Foundation \cite{grant}. We thank the authors of Ref.~\cite{Giha2025} for kindly providing the experimental data. A. Rahmatinejad gratefully acknowledges the generous support of F. Gorji and H. Tofighitehranimonfared during the course of this research.

\end{document}